\documentclass[aps,prb,twocolumn,floatfix,superscriptaddress]{revtex4}
\usepackage{amsmath}
\usepackage{amssymb}
\usepackage{graphicx}
\usepackage{dcolumn}
\usepackage{bm}

\begin{document}

\title{Spin-droplets in confined quantum Hall systems}

\author{E. R{\"a}s{\"a}nen}
\email[Electronic address:\;]{esa@physik.fu-berlin.de}
\affiliation{Institut f{\"u}r Theoretische Physik,  
Freie Universit{\"a}t Berlin, Arnimallee 14, D-14195 Berlin, Germany}
\affiliation{European Theoretical Spectroscopy Facility (ETSF)}
\author{H. Saarikoski}
\affiliation{Kavli Institute of NanoScience, Delft University
of Technology, 2628 CJ Delft, the Netherlands}
\author{A. Harju}
\affiliation{Laboratory of Physics, Helsinki University of Technology,
P.O. Box 4100, FI-02015 HUT, Finland}
\author{M. Ciorga}
\affiliation{Institut f{\"u}r Experimentelle und Angewandte Physik,
Universit{\"a}t Regensburg, Universit{\"a}tsstra\ss e 31, Regensburg 93040,
Germany}
\author{A. S. Sachrajda}
\affiliation{Institute for Microstructural Sciences,
National Research Council of Canada, Ottawa, Canada K1A 0R6}

\begin{abstract}
Two-dimensional semiconductor quantum dots are studied 
in the the filling-factor range $2<\nu<3$. We find both
theoretical and experimental evidence of a collective 
many-body phenomenon, where a fraction of the trapped 
electrons form an incompressible spin-droplet on
the highest occupied Landau level. The phenomenon
occurs only when the number of electrons in the quantum 
dot is larger than $\sim 30$. We find the onset of the 
spin-droplet regime at $\nu=5/2$. This proposes a 
finite-geometry alternative to the Moore-Read-type 
Pfaffian state of the bulk two-dimensional electron gas.
Hence, the spin-droplet formation may be related to the observed 
fragility of the $\nu=5/2$ quantum Hall state in narrow 
quantum point contacts. 
\end{abstract}

\maketitle


The fractional quantum Hall effect at filling factor 
$\nu=5/2$ has recently attracted both experimental and
theoretical interest. The $\nu=5/2$ state in the 
two-dimensional electron gas (2DEG)~\cite{willett} has commonly 
been interpreted as a condensate of paired fermions
described by a Pfaffian wave function.~\cite{moore}
Alternative models that capture the incompressible
nature of this state have been proposed using the
composite fermion theory.~\cite{toke} 
The non-Abelian braiding statistics involved in the
excitations of the $\nu=5/2$ state makes it a possible 
candidate for topological quantum computing with 
high error tolerance.~\cite{sarma} In experiments, 
however, the state has been found to 
be fragile, and it may break down in the presence of
impurities, or in narrow quantum point 
contacts.~\cite{miller}

Finite-size counterparts of both integer and fractional 
quantum Hall states have been observed and characterized 
in confined 2D systems such as in semiconductor quantum 
dots (QDs).~\cite{reimannreview} Due to the external 
confinement, these many-body states often differ 
considerably from the corresponding states in the 2DEG.
In a recent theoretical study,~\cite{half} the QD 
counterpart of the $\nu=5/2$ state was defined,
but the half-filled Landau-level was shown to be
poorly described by the Pfaffian.

In this work we show that in QDs
the ground states in the $5/2\geq \nu >2$ quantum Hall
  regime are characterized by fragmentation of the
  spin and charge densities into integer filling
  factor domains. This results in the formation of {\em spin-droplets}
  (SDs), incompressible droplets of spin-polarized
  electrons in the highest occupied Landau level.
We find that the SD formation is a collective
phenomenon interacting electrons, 
and it occurs only when the total number of electrons in the
QD is sufficiently large ($N \gtrsim 30$). 
We detect signatures of the SD formation
in two distinctive quantum-transport experiments in a very good agreement
with our theoretical results. The emergence of SDs is in
contrast with the previous understanding which assumed 
ground-state oscillations with a low spin polarization.~\cite{klitzing01,tarucha00}
As a finite-geometry alternative to the Pfaffian state,
the formation of SDs may explain the fragility of 
the $\nu=5/2$ state in narrow quantum point contacts 
in the 2DEG. The results are therefore
relevant for the proposed experiments to probe the non-Abelian
characteristics of the $\nu=5/2$ quantum Hall state, since they make
use of confined geometries.~\cite{sternbondersonfeldman}

We analyze many-electron ground states of QDs 
in magnetic fields both theoretically and experimentally. 
In our theoretical model 
we use the standard effective-mass Hamiltonian
\begin{equation}
H=\sum^N_{i=1}\left[
 \frac{({\bf p}_i+e {\bf A} )^2}{2 m^*}
+V_{{\rm c}}(r_i)+g^*\mu_B S_{z,i}\right] + \frac{e^2}{4\pi \epsilon} \sum_{i<j}
\frac{1}{r_{ij}},
\label{hamiltonian}
\end{equation}
where $V_{{\rm c}}(r)=m^*\omega_0^2 r^2/2$ is the
external confinement, $m^*=0.067m_e$, $\epsilon=12.7\epsilon_0$, and
$g^*=-0.44$ are the effective material parameters for GaAs semiconductor
medium, and ${\bf A}$ is the vector potential of the homogeneous 
magnetic field ${\bf B}$ perpendicular to the QD plane. 
The many-electron problem is solved computationally using the 
(spin-)density-functional theory (DFT) and the
variational quantum Monte Carlo method (QMC). 
Numerical details of the methods in QD applications
are given in Refs.~\onlinecite{pss} and \onlinecite{ari_vmc}, 
respectively.

\begin{figure}
\includegraphics[width=0.75\columnwidth]{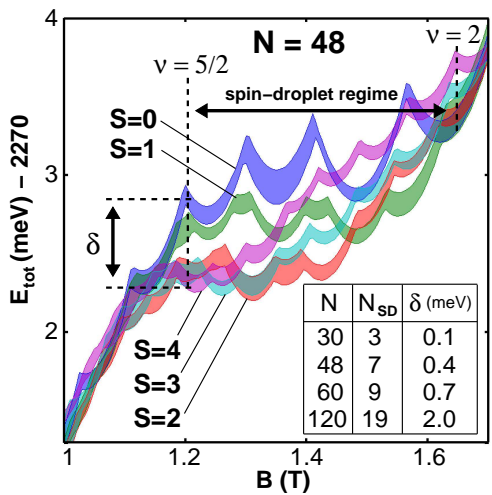}
\caption{
(Color online)
Total energies of spin states of the $N=48$ quantum dot ($\hbar\omega_0=2$ meV)
calculated with the quantum Monte Carlo method in the spin-droplet
regime. The line widths denote the statistical error in the calculations.
The table shows the density-functional-theory result for the maximum number 
of electrons in the spin-droplet 
$N_{\rm SD}$ as a function of $N$, as well as the total-energy decrease
at $\nu=5/2$ due to the spin polarization.
}
\label{fig1}
\end{figure}
\begin{figure}
\includegraphics[width=1\columnwidth]{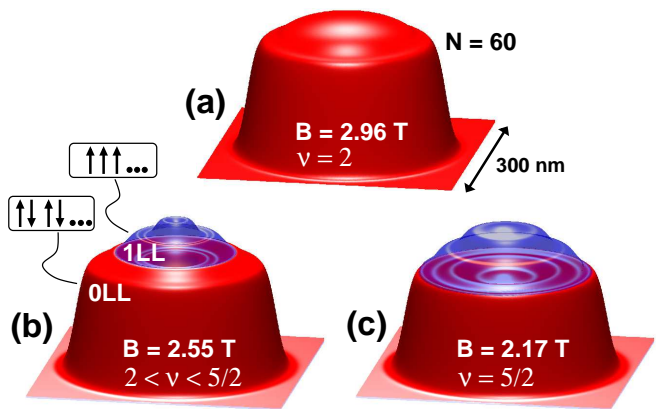}
\caption{
(Color online)
Electron densities of a 60-electron
quantum dot calculated with the density-functional method
at $\nu=2$ (a), at an intermediate state
between $\nu=2$ and $\nu=5/2$ (b), and at $\nu=5/2$ (c). 
The red (solid) and blue (transparent)
regions mark the electron densities
in the spin-compensated lowest (0LL) and the spin-polarized
second-lowest Landau levels (1LL), respectively. The series of figures thus
demonstrate the formation and growth of the spin-droplet in
the second lowest Landau level.
}
\label{fig2}
\end{figure}

Figure~\ref{fig1} 
shows the total energies of different spin states of a
48-electron QD calculated with QMC.
The DFT method shows qualitatively the same behavior.
Here we point out that in QDs the filling factor in the $\nu\geq 1$
regime can be approximated by a formula $\nu=2N/N_{\rm 0LL}$, 
where $N_{\rm 0LL}$ is the number of electrons in the lowest
landau level (0LL).~\cite{half}
At magnetic fields $B<~ 1.1$ T the degeneracy of the 
many-electron ground states with different spin polarization is high.
When $B$ is increased, the partially spin-polarized states 
become lower in energy with respect to the $S=0$ state.
Finally at $B\sim 1.2$ T the QD reaches the highest spin polarization ($S=4$).
This state can be identified as the $\nu=5/2$ state, which is broken into
two incompressible integer filling-factor domains,
$\nu=2$ and $\nu=3$, inside the QD.~\cite{half}
The electron occupations are separated into the spin-compensated
lowest 0LL, forming a relatively flat background of electrons,
and into the {\em totally spin-polarized} second-lowest Landau level (1LL),
forming a SD. The SD gradually diminishes at $1.2 <~ B <~ 1.7$ T
corresponding to $5/2 \geq \nu > 2$, which we define as the 
{\em SD regime}. The total electron densities and the
respective densities of the 0LL and 1LL in the SD regime
are visualized in Fig.~\ref{fig2}. 
Both the DFT and QMC results show the same topology.

We define the total-energy decrease achieved from the 
spin polarization of the 1LL as 
$\delta={\rm max}[E(S=0)-E(S_{\rm max})]$ calculated
at $\nu=5/2$, i.e., at the onset of the SD regime.
The Zeeman energy has only a minor $(<10\%)$ contribution 
in $\delta$ in these magnetic fields,
the rest being due to electron-electron interactions.
Our calculations show that SDs do not form below 
$N \lesssim 30$ due to the loss of spin-polarization 
of the 1LL. The table in Fig.~\ref{fig1} demonstrates that
the size of the SD $N_{\rm SD}$ grows with $N$, and
$\delta$ is roughly linearly proportional to $N-30$.
This indicates high stability of SDs in large QDs
approaching the 
width of the narrow quantum point contacts
made in 2DEG.~\cite{miller}
The DFT method predicts that 
the SD formation is insensitive to confinement strength
at least for confinements $1 \ldots 4$ meV.

The spin-compensated DFT calculations in the SD regime 
indicate a high degeneracy of
single-particle (Kohn-Sham) states of the 1LL 
electrons near the Fermi level. The
SD formation can then be understood in terms of the Stoner
criterion which states that in the presence of correlations between
electrons of the same spin and high density of states near the Fermi
level, the system prefers ferromagnetic alignment which reduces the
degeneracy.~\cite{stoner} We note that the reduction in degeneracy from
what is found in the 2DEG is due to the external confining potential. 

We discuss now the evidence for the emergence 
of SDs in the electron transport data
in two experimental setups for {\em lateral} 
and {\em vertical} QD devices, respectively.
The emergence of finite-size counterparts of integer
and fractional quantum Hall states
is reflected on the energetics of the QD system.
Existence and properties of these states can be probed
by measuring, e.g., the chemical potential
$\mu(N,B)=E_{\rm tot}(N,B)-E_{\rm tot}(N-1,B)$,
which is the energy required by the $N$th electron to enter the QD.
Several experimental methods have been developed for this purpose
including Coulomb blockade,~\cite{mceuen}
capacitance,~\cite{ashoori} and charge detection 
techniques.~\cite{field}

Our setup for lateral QDs combines Coulomb and
spin-blockade measurements described in detail in 
Ref.~\onlinecite{ciorga00}.
Two samples (A and B) of lateral devices
were manufactured, and
standard low-power ac measurement techniques were used: 
a small ac voltage with an amplitude $dV=10\,{\rm \mu\rm V}$ and frequency
$f=23$ Hz was applied across the sample, and a current amplifier 
and lock-in amplifier were used to measure the resulting 
spin-resolved current up to $N=48$.~\cite{ciorga00} The bulk mobility
of the Al$_{\rm x}$Ga$_{\rm 1-x}$As/GaAs wafer used
was $2\times10^6$ cm$^2$/(Vs).
Another set of data was obtained from
the electron transport experiments of a vertical QD device
by Oosterkamp and co-workers,~\cite{oosterkamp} who measured the
Coulomb-blockade oscillations up to $N=39$.

\begin{figure*}
\includegraphics[width=0.8\linewidth]{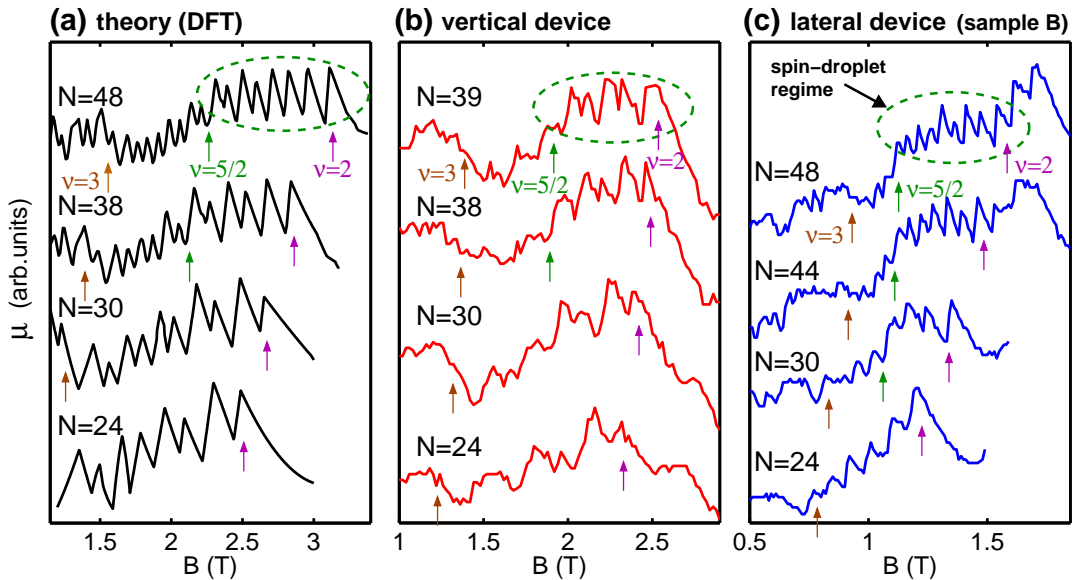}
      \caption{(Color online)
Chemical potentials calculated with density-functional theory 
(a) and measured
from a vertical (b) and lateral (c) quantum-dot devices
for various electron numbers. Both experiments show
the emergence of the spin-droplet signals in the peak position data
when $N\gtrsim 30$ in agreement with the theoretical result.
Data for the vertical device in (b) is courtesy of
L. Kouwenhoven, Delft.~\cite{oosterkamp}
}
\label{fig3}
\end{figure*}

Figure~\ref{fig3}
shows the DFT results for chemical potentials of
$N=24\ldots 48$ in comparison with the experiments.
The confinement strength ranged from 2
to 4 meV depending on 
the electron number.
The oscillations in $\mu$ correspond to ground-state 
transitions. Overall, we find an 
excellent qualitative agreement between the theory and experiments. 
At $\nu=5/2$ in high electron numbers there is a
step feature followed by a plateau region 
superimposed on the oscillations.
Our calculations show that the growing of the polarization energy $\delta$
with $N$ is crucial for the formation of these features. 
We identify them as signatures of the SD formation.
In both the experimental data shown in Figs.~\ref{fig3}(b) and (c)
we note a gradual decay of the SD plateau
at low electron numbers. This is in agreement with the theoretical
data in Fig.~\ref{fig3}(a). The decay is due to the
fact that the spin polarization of the 1LL is lost when
$N\lesssim 30$ (see the table in Fig.~\ref{fig1}).

\begin{figure}
\includegraphics[width=0.9\columnwidth]{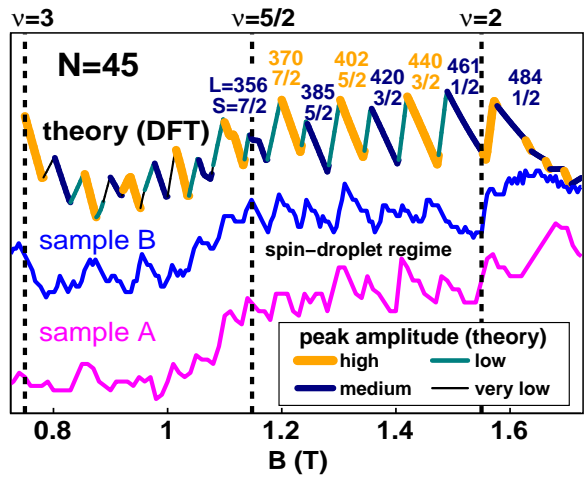}
\caption{
(Color online)
Chemical potential of two samples of lateral 45-electron dots obtained using 
spin-blockade spectroscopy, and the corresponding theoretical result from 
density-functional calculations. Both samples show a clear spin-droplet plateau
at $5/2\geq\nu>2$ in agreement with the theory.
The high, medium, low, and very low amplitude peaks in
theory correspond to transport involving the 0LL spin $\uparrow$, 0LL spin
$\downarrow$, 1LL spin $\uparrow$, and 1LL spin $\downarrow$ states,
respectively. The numbers correspond to the {\em z}
components of the ground-state total angular momenta (L) and spins (S)
for the high and medium-amplitude states.
}
\label{fig4}
\end{figure}

In Fig.~\ref{fig4} we
compare the DFT result of the chemical potentials for $N=45$
to two samples of the lateral QDs. In the SD
regime the agreement between the theory and experimental
data is very good: the oscillations in $\mu$ match
almost peak-by-peak which is an indication of discrete transitions
in the many-body ground state. The corresponding
values for the ground-state total angular momenta (L) and spins (S) 
({\em z} components) calculated with the DFT are marked in Fig. \ref{fig4}.
The similarity of the results from 
two different QD samples confirms our prediction of the stability 
of the SD even in the presence of inevitable impurities
in the samples.

The electrons enter the lateral QDs from spin-polarized magnetic edge
states of a 2DEG through tunneling barriers. Therefore current through the
dot is enhanced if the transport involves a single-particle state in the
QD which has spin parallel to the spin in the contacts. As a
consequence the amplitude pattern shows a characteristic chessboard
pattern of high and medium conductivity regions.~\cite{klitzing01,tarucha00}
The conductivity pattern obtained from the DFT calculations is
shown in Fig.~\ref{fig4} and agrees with the
published amplitude data.~\cite{ciorga00}
The cusp in the peak position data of the lateral devices at $B=1.7\;{\rm T}$
is related to the chemical potential jumps in the 2DEG.~\cite{ciorga00}

Our theoretical and experimental results suggest that the SD formation
could be the origin of the previously observed fragility of the bulk
$\nu=5/2$ state in, e.g., the vicinity of narrow quantum point
contacts.~\cite{miller} In our scenario, confinement effects such as
point-contact boundaries or impurities could locally form
SDs with distinct phase boundaries. The formation of
SD-like regions would be analogous to incompressible regions which appear
in some theories of the quantum Hall effect
in a disordered 2DEG.~\cite{chklovskii}
This type of behavior
has been observed in the 2DEG by Tessmer and co-workers~\cite{tessmer}
who, using scanning-probe imaging techniques, observed
the formation of a system of incompressible droplets with reduced electron 
density in a perturbed quantum Hall liquid.
In another study Finkelstein and co-workers~\cite{finkelstein} 
used the same probing technique to map the random potential 
believed to be responsible for the localization of electrons 
near integer quantum Hall plateaus. We point out that additional 
theoretical studies, where the Pfaffian $\nu=5/2$ state is 
exposed to confinement, are needed to confirm this breakdown mechanism.
Direct experimental evidence of the SD
formation in QDs may be attained using accurate imaging methods.~\cite{fallahi}

We finally point out that the large spin polarization of the SD state
is encouraging for potential applications in the field of spintronics.
Spin-polarized quantum Hall states such as $\nu=1$ and $\nu=1/3$
have been recently proposed as building blocks for spintronics 
devices through the spin-orbit coupling tuned by an external 
electric field.~\cite{califano} Spin-polarized electrons in SDs
could be exploited in similar applications. 


To conclude, we have shown evidence for spin-droplet formation
in quantum dots at large electron numbers. The spin-droplets
are found to be largest at magnetic fields corresponding to $\nu=5/2$.
The computational results given by two different 
many-electron methods agree with experimental sets of data
from electron transport measurements, and show that the spin-droplet
formation is a collective many-electron phenomenon.
The formation of locally confined spin-droplet states in the
presence of inhomogeneities such as impurities
or point contacts may explain the observed fragility of the 
$\nu=5/2$ quantum Hall state in the 2DEG.
Our findings stress the importance of confinement 
effects on quantum many-body states in attempts to realize 
topological quantum computing.

We thank L. Kouwenhoven for the kind permission to 
reproduce their experimental data for the vertical 
quantum-dot device, and P. Zawadzki, 
Z. Wasilewski, and P. Hawrylak for assistance and 
discussions concerning the lateral dot experiments.
This work was supported
by the EU's Sixth Framework Programme through 
the Nanoquanta Network of Excellence (NMP4-CT-2004-500198),
the Academy of Finland, the Finnish Academy of Science and Letters
through the Viljo, Yrj{\"o} and Kalle V{\"a}is{\"a}l{\"a}
Foundation, and the Canadian Institute for Advanced Research.

\end{document}